\documentclass[aps,prb,twocolumn,showkeys]{revtex4-2}
\usepackage{amsmath}
\usepackage{graphicx}
\begin{document}
\title{Beyond the Drude model: surface and non-local effects in near-field radiative heat transfer and the Casimir puzzle}

\author{Jian-Sheng Wang}
\affiliation{Department of Physics, National University of Singapore, Singapore 117551, Republic of Singapore}

\date{15 February 2025}

\begin{abstract}
We study the charge and current response functions $P$ and $\Pi$ in a semi-infinite metal block using the
electron surface Green's functions. The surface electrons behave similarly to a two-dimensional Fermi gas but are 
strongly damped due to coupling to the bulk.  This substantially reduces the region of validity of the Drude model for $P$, which requires
the frequency $\omega \gg \max( v_F q, 1/\tau)$, here $v_F$ is the Fermi velocity, $q$ is the wavevector and $\tau$ is 
an effective relaxation time.  As a consequence, for typical metal in near-field heat transfer, 
the Coulomb interaction goes as $1/d^4$ with the distance of the vacuum gap instead of the well-known $1/d^2$ of
Drude model result.  The current response $\Pi$ is shown to be highly anisotropic.  The Drude model describes well  
the transverse directions parallel to the surface but is very different in the normal direction up to about 
100 lattice sites away from the surface.   These ideas and the residue diamagnetic effect of a nonzero $\Pi$ on the surface at zero 
frequency still cannot resolve the Casimir puzzle. 
\end{abstract}


\maketitle

\section{Introduction}

Heat transfer between parallel plates has been an archetype problem with the pioneering work by Polder and van Hove \cite{Polder71}
using Maxwell's equations and fluctuation-dissipation theorem, known as the fluctuational electrodynamics \cite{Rytov53,Rytov89,Volokitin07,Basu09,Song15,Biehs21,Messina17}.  Many experiments 
\cite{Hargreaves69,Domoto70,Kittel05,Ottens11}
have confirmed the near-field enhancement of heat transfer compared to the 
black-body results by a thousand-fold at the nanometer scale \cite{Kim15,Cui17,Kloppstech17}.  It comes to us as a surprise that there are still unexplored
parameter regimes in which the PvH theory may be unsuitable.   This leads us to examine the validity of 
the Drude model, a standard reference system for metal systems.   

Drude model has been of paramount importance in describing metals for their transport properties and near-field
heat transfer and Casimir forces \cite{Casimir48,Messina15}.  However, there might be cases where the picture of Drude's transport may be broken down.
Here, we show by analytic consideration, as well as numerical simulation using a cubic lattice model, that this is indeed the
case that the continuum theory of Polder and van Hove (PvH) cannot correctly describe the longitudinal part of the 
physics which is controlled by the Coulomb interaction.   The PvH  theory based on Drude model conductivity predicts 
$1/d^2$ distance dependence, while the actual simulation result is $1/d^4$.   We can explain this $1/d^4$ scaling
by revising the scalar field polarizability as given by surface charge correlation, which leads to $P \sim a + i b\omega$, where
$a$ and $b$ are some constants.  On the other hand, the Drude model version is $P \sim q^2/( \omega(\omega + 2 i \eta))$. 

The Casimir puzzle has been around for nearly 20 years and is still unresolved
\cite{Mostepanenko06,Klimchitskaya22}.   This concerns the following:
The Drude model gives a poor fit to the experimental data at the thermal regime at a distance of about micrometers.  
However, if the dissipation is set to zero so that the Drude model becomes the plasma model, it fits high-precision
data.  This is puzzling, as dissipation must exist at room temperature, and the reason for its absence is not fully understood.    We also look into this but without success.
However, our numerical observation on the fully non-local polarizability $\Pi$ for the surface, which depends on the
coordinates $(z,z')$, distances away from the surface, may be the key for future investigation. 

The following sections describe the local Drude model and its nonlocal generalization.  We then study the
heat transfer between metal plates and discuss some still unsolved issues when the metal is treated as a cubic
lattice model.  In particular, we discuss a possible resolution of the Drude vs.\ plasma puzzle in thermal Casimir force in
view of the nonlocal polarizability $\Pi$. 
We conclude in the last section. 

\section{Drude model}

The current produced by a static electric field in metal is well-described by Ohm's law through the Drude model \cite{PBAllen},
\begin{equation}
{\bf j} = \sigma {\bf E}
\end{equation}
with
\begin{equation}
\sigma = e^2 \left( \frac{n}{m} \right) \tau.
\end{equation}
It is useful to think of $n/m$ as a single quantity when, e.g., the mass $m$ is not well-defined, e.g., in graphene.
In Boltzmann's transport theory, it has the expression 
\begin{equation}
\frac{n}{m} = 2 \int \frac{d^{3} {\bf k}}{(2\pi)^3} 
\frac{1}{3} {{\bf v}_{\bf k}}^2 \left( - \frac{\partial f }{\partial \epsilon_{\bf k}} \right),
\end{equation}
where 2 is for spin degeneracy,  $f$ is the Fermi function, $\epsilon_{\bf k}$ is the electron dispersion relation, ${\bf v}_{\bf k}$ is
the group velocity, and the integral is over the first Brillouin zone. 

For an arbitrary time-dependent electric field, we can revise the formula as, in the time domain
\begin{equation}
\label{eqsigmat}
\sigma(t) = \theta(t) e^2 \frac{n}{m} e^{-t/\tau}.
\end{equation}
This gives the current response as
\begin{equation}
\label{eqjtEt}
 {\bf j}(t) = \int^t dt' \sigma(t-t') {\bf E(t')}.
\end{equation}
It represents a simple exponential relaxation, which we can obtain by solving Newton's equation of motion 
with a damping,
\begin{equation}
m \frac{d\bf v}{dt} = - m \frac{ {\bf v}}{\tau} + (-e) {\bf E}(t).
\end{equation}
 When $\tau \to \infty$, we can call it a diamagnetic response or plasma response, as then, we can write the
current response as
\begin{equation}
{\bf j}(t) = (-e) n {\bf v}(t) =  - \frac{e^2 n}{m} {\bf A}(t).
\end{equation}
Here ${\bf A}(t) = - \int^t dt' {\bf E}(t')$ is the vector potential (in $\phi = 0$ gauge).
Fourier transforming Eq.(\ref{eqsigmat}) into the frequency domain, we have
\begin{equation}
\sigma(\omega) = \frac{e^2 n}{m} \frac{i}{ \omega + i /\tau}.
\end{equation}
As for any response functions in Fourier space, it is analytic in the upper half plane of complex $\omega$.

\section{Drude model $\Pi$}
The response function $\Pi$, or polarizability, is defined by 
\begin{equation}
{\bf j}(t) = - \int dt' \Pi(t-t') {\bf A}(t').
\end{equation}
Using Eq.~(\ref{eqjtEt}) and ${\bf E} = - \partial {\bf A}/{\partial t}$, performing an integration by parts, we find
\begin{equation}
\frac{d \sigma(t)}{dt} = \Pi(t).
\end{equation}
So, in the time domain, it is
\begin{equation}
\Pi(t) = \delta(t) \frac{e^2n}{m} - \theta(t) \frac{e^2 n}{m \tau} e^{-t/\tau}.
\end{equation}
We can call the first term ``diamagnetic'' and the second term RPA (random phase approximation). 
In the frequency domain, it is 
\begin{align}
\label{eq-local-Drude-Pi}
\Pi(\omega) & = (-i\omega) \sigma(\omega) = -\epsilon_0 \omega^2 (\epsilon-1) \nonumber \\
&= \frac{e^2 n}{m} \frac{\omega}{ \omega + i /\tau}.
\end{align}

The above discussion is based on a local theory, i.e., wave-vector $\bf q$ independent.  Taking into account
the wavevector dependence, we can express $\Pi$ per site on a square or cubic lattice as, through the current-current correlation \cite{Mahan00,Pablo25},
\begin{align}
\label{eq-RPA-Pi}
\Pi_{\alpha\beta}^{{\rm RPA}}(\omega, {\bf q}) &= 
\frac{2e^2}{N} \sum_{{\bf k}}  v^\alpha({\bf k},{\bf q}) v^\beta({\bf k}, {\bf q}) \times \qquad \nonumber \\
 &\qquad \frac{ f(\epsilon_{\bf k} + i \eta) - f(\epsilon_{{\bf k}+{\bf q}} - i\eta)
}{ \hbar \omega + 2 i \eta + \epsilon_{\bf k} - \epsilon_{{\bf k} + {\bf q}} }.
\end{align}
Here and below, the Greek indices denote the Cartesian directions, $x$, $y$, or $z$, and not 4-vector indices.  The expression is valid for one-band model with ${\bf v}({\bf k},  {\bf q}) = \frac{1}{2} ( {\bf v}_{\bf k} + {\bf v}_{{\bf k} + {\bf q}})$ \cite{WenQFT}.  The $\alpha$ component of $\bf v$ is
denoted by $v^\alpha$.
$N$ is the number of lattice sites (or the number of $\bf k$ points). 
The imaginary argument in the Fermi functions is due to the application of the residue theorem to the energy integral of $g^r g^< + g^< g^a$
(see Eq.~(\ref{eq-RPA-Eint}) below or \cite{Haug08}), where $g$ is the electron
Green's function.  The total is obtained by adding a constant diamagnetic term, which we can use the gauge invariance to fix to be
the negative of the RPA result evaluated at zero frequency and wavevector:
\begin{equation}
\label{eq-gauge-0}
\Pi(\omega, {\bf q}) = \Pi^{{\rm RPA}}(\omega, {\bf q})  - \Pi^{\rm RPA}(0, {\bf 0}).
\end{equation}
Evaluating Eq.~(\ref{eq-RPA-Pi}) at ${\bf q} = 0$, we reproduce the Drude model result, with the identification of $\hbar/\tau = 2 \eta$, here
$\eta$ is the electron relaxation parameter in the retarded Green's function (for one mode) as $g^r(E) = 1/( E + i\eta - \epsilon_{\bf k})$.

\section{$\Pi$ in a semi-infinite block}
To study the Casimir effect or heat radiative transfer between two semi-infinite metal blocks modeled as cubic lattice, we also need a version
of $\Pi$ consistent with the geometry, i.e., periodic in the $x$ and $y$ direction, but an open boundary condition
in $z$ direction, say at $z = ja$, $j = 1$, 2, $\cdots$, $\infty$.   This means that at the sites $j=1$, the electrons can only
hop back, or on the surface, but not to $j=0$. This can be obtained by the surface Green's function in the cubic
lattice as \cite{Reuter10} 
\begin{equation}
\label{eq-grjj}
g^r(E, {\bf q}_{\perp}, j, j') = \frac{  \lambda^{|j-j'|}  -  \lambda^{j+j'}   }{t (\lambda - 1/\lambda)},
\end{equation}
where $E$ is energy (Fourier transform of the time), ${\bf q}_{\perp} =(q_x, q_y)$ is the wavevector in the transverse
direction,  $t = \hbar^2/(2ma^2)$ is the hopping parameter.  
The complex number $\lambda$ is determined by the quadratic equation
\begin{equation}
t \frac{1}{\lambda} + E + i \eta - \epsilon_{{\bf q}_\perp} + t \lambda = 0, \quad |\lambda| < 1.  
\end{equation}
Here $\epsilon_{{\bf q}_\perp} = -2 t \bigl( \cos(q_x a) + \cos(q_y a) \bigr)$ is the dispersion relation of the electron
on the plane.  
In Eq.~(\ref{eq-grjj}), the first term is the bulk contribution; it is the usual Green's
function, which represents a traveling wave moving away from the source. 
The second term is the surface contribution, which decays
to 0 deep into the bulk.    Due to the boundary at $z=0$, the interference effect develops.   The Green's function is nonlocal 
in $z$.    

The dissipation of the electron is taken care of in two ways: by a wide-band approximation of the electron self-energy through $\eta$, 
and because it is an open system, waves propagating to infinity will never reflect back.   Compared to the 
actual 2D Green's function of $1/(E+i\eta - \epsilon_{{\bf q}_\perp})$,  the surface Green's function is much more damped due to the self-energy 
from the bulk. 
In terms of the surface Green's function $g$, the polarizability  $\Pi$ can be expressed, in the $x,y$ sector, as
\begin{align}
\Pi_{\alpha\beta}^{\rm RPA}(\omega, {\bf q}_\perp, j, j')  = - \frac{2i e^2}{L^2} \sum_{{\bf p}_\perp} 
v^\alpha({\bf p}_\perp, {\bf q}_\perp) v^\beta({\bf p}_\perp, {\bf q}_\perp)  \quad \nonumber \\
 \int \frac{dE}{2\pi} \Big[ g^r(E, {\bf p}_{\perp} + {\bf q}_\perp, j, j') g^<(E - \hbar \omega, {\bf p}_{\perp}, j', j) \qquad \nonumber \\
+   g^<(E, {\bf p}_{\perp} + {\bf q}_\perp, j, j') g^a(E - \hbar \omega, {\bf p}_{\perp}, j', j) \Big] \quad
\label{eq-RPA-Eint}
\end{align}
Here $\alpha$ or $\beta$ takes $x$ and $y$ only.  The discrete index $j$ is related to the $z$ coordinate by
$z=aj$.  The cross-section of the plate has $L \times L$ sites.  The lesser Green's function is obtained by the fluctuation-dissipation theorem,
$g^< = -f(E) (g^r - g^a)$, and $g^a = (g^r)^\dagger$.   The scalar $P$ for the charge-charge
correlation is obtained if we set the velocities in the above expression to 1.
The expressions involving the $z$ component are more complicated.   We
refer to  arXiv:1607.02840 for details \cite{Peng1607arxiv}.

The important point is the diamagnetic term \cite{Wang-Mauro-24}.   We derive the diamagnetic term starting from the Peierls' substitution 
Hamiltonian \cite{Peierls33} defined on a cubic lattice.
\begin{equation}
\hat{H}({\bf A}) = -t \sum_{{\bf l}, {\bf a}} c_{{\bf l} + {\bf a}}^\dagger c_{\bf l} \exp\left( - \frac{ie}{\hbar}  \int_{\bf l}^{{\bf l} + {\bf a}} \!\!\! {\bf A} \cdot d{\bf r} \right), 
\end{equation}
here $\bf l$ runs over the lattice sites, $\bf a$ is a vector pointing to one of the six neighbors of a given site, and $(-e)$ is the electron charge.
When $z=ja\leq  0$ which does not exist, we can conveniently set the corresponding physical quantities to 0. 
We approximate the line integral by a straight line and using the trapezoidal integration rule:
\begin{equation}
\int_{\bf l}^{{\bf l} + {\bf a}} \!\!\! {\bf A} \cdot d{\bf r} = \frac{1}{2} \left( {\bf A}_{\bf l} + {\bf A}_{{\bf l}+{\bf a}} \right) \cdot {\bf a}. 
\end{equation}
A variational derivative with respect to $\bf A$ can identify the electric current operator, as 
\begin{equation}
\delta \hat{H} = - \sum_{\bf l} {\bf  I}_{\bf l} \cdot \delta {\bf A}_{\bf l}.
\end{equation}
From the Hamiltonian, we can identify the current as
\begin{align}
{\bf I}_{\bf l} = & -  \frac{iet}{2\hbar} \sum_{\bf a} {\bf a} \, (  c_{{\bf l} + {\bf a}}^\dagger c_{\bf l} +  c_{\bf l}^\dagger c_{{\bf l} - {\bf a}}  ) \nonumber \\
& -   \frac{e^2 t}{4\hbar^2}\sum_{\bf a} \, {\bf a} {\bf a} \cdot \Big[ \bigl({\bf A}_{{\bf l} + {\bf a}} + {\bf A}_{\bf l} \bigr) c_{{\bf l} + {\bf a}}^\dagger c_{\bf l}  \nonumber \\
& \qquad\qquad \bigl( {\bf A}_{\bf l} + {\bf A}_{{\bf l}- {\bf a}} \bigr)  c_{\bf l}^\dagger c_{{\bf l} - {\bf a}}
  \Bigr].
\end{align}
The $\bf aa$ factor should be interpreted as a dyadic dotted into the vector $\bf A$.   The first term independent of the vector potential is the
paramagnetic current; the rest of the terms proportional to the vector potential is the diamagnetic current.  If we define
\begin{equation}
{\bf I}_{\bf l} = {\bf I}_{\bf l}^{\rm para}  - \sum_{{\bf l}'} \Pi^{\rm dia}_{{\bf l}{\bf l}'} \cdot {\bf A}_{{\bf l}'},
\end{equation}
The tensor in space index $\Pi^{\rm dia}$ can be identified.  Omitting some details, taking the expectation value of the operators to 
get $g^<$, we can express the diamagnetic tensor in Fourier space in transverse direction and real site in $z$ direction as
\begin{align}
\Pi^{\rm dia}_{\alpha\alpha}(\omega, {\bf q}_\perp, j, j') = - \delta_{j, j'} \frac{2ie^2}{m} \frac{1}{L^2} \times  \qquad\qquad\qquad\qquad \nonumber \\
\qquad \sum_{{\bf p}_\perp} 
 \int \frac{dE}{2\pi} g^<(E, {\bf p}_\perp, j, j) \cos(p_\alpha a) \frac{1 + \cos(q_\alpha a)}{2} ,
\end{align}
here $\alpha  = x$ or $y$ only.   In the $xy$ sector, $\Pi$ is diagonal in the sites.  The $z$ direction is 
\begin{align}
\Pi^{\rm dia}_{zz}(\omega, {\bf q}_\perp, j, j) =  - \frac{ie^2}{4m} \frac{1}{L^2}\int \frac{dE}{2\pi}    \qquad\qquad\qquad\qquad \nonumber \\
\qquad \sum_{{\bf p}_\perp} \Big[ 
 g^<(E, {\bf p}_\perp, j+1, j) + g^<(E, {\bf p}_\perp, j, j-1) \qquad\qquad \nonumber \\
 \quad + \, g^<(E, {\bf p}_\perp, j-1, j) + g^<(E, {\bf p}_\perp, j, j+1) \Big]. \qquad \qquad
\end{align}
The off-diagonal terms are nonzero when $|j - j'| = 1$, 
\begin{align}
\Pi^{\rm dia}_{zz}(\omega, {\bf q}_\perp, j, j\pm1) =  - \frac{ie^2}{4m} \frac{1}{L^2}\int \frac{dE}{2\pi}    \qquad\qquad\qquad\qquad \nonumber \\
\qquad \sum_{{\bf p}_\perp} \Big[ 
 g^<(E, {\bf p}_\perp, j\pm 1, j) + g^<(E, {\bf p}_\perp, j, j\pm 1) \Big] . \qquad\qquad 
\end{align}

\subsection{Choice of model parameters for Gold}
We set parameters to match the properties of gold \cite{Ashcroft-Mermin76}. 
The electron density $n = 1/a^3 =  5.90\,\times10^{22}$cm$^{-3}$, this gives a lattice constant 
$a = 4.856\,$a.u.~(2.56\AA).  The abbreviation a.u.~here and below stands for Hartree atomic
units with $e=\hbar=m_e=1$ and $4\pi\epsilon_0 = 1$.  
 Each site is half-filled with one electron at the chemical potential $\mu = 0$. 
 At half filling, the carrier density is $n = 1/a^3$, giving the radius $r_s = 3.01\,$bohr.
The plasma frequency determines the hopping parameter $t=0.0635\,$a.u.~(1.73 eV).  
This gives $n/m$ and the plasma frequency 
\begin{equation}
\frac{a^3n}{m} \approx 1.0\,{\rm a.u.},\quad \omega_p = \sqrt{\frac{ne^2}{m \epsilon_0}} \approx 9.02\, {\rm eV}.
\end{equation}
The Fermi velocity averaged over the Fermi surface is 
$v_F = 0.7 \times 2at/\hbar = 0.217 \,$a.u.   We set the damping parameter $\eta = 10^{-3}$a.u.~(0.027 eV),
temperature $T = 300\,$K. 

Some of the length scales involved are the mean-free path $l$, 
\begin{equation}
l = v_F \tau = 109 \,{\rm a.u.\ } (22\,a).
\end{equation}
Here the relaxation time $\tau = \hbar /(2 \eta) =500\,{\rm a.u.}$. 
The DC conductivity is 
\begin{equation}
\sigma = \frac{e^2 n \tau}{m} = 4.4\, {\rm a.u.}
\end{equation} 
The two length scales, the London penetration depth and skin depth (at $\hbar\omega = k_BT$), are
\begin{align}
\delta_L = & \frac{c}{\omega_p} = 414\,{\rm a.u\ } (85\, a), \\
\delta_T = & \sqrt{\frac{\hbar}{\mu_0 \sigma k_B T}} \approx 582\,{\rm a.u.\ } (120\, a).
\end{align} 
These length scale parameters are important with respect to our system size parameter,
$L_z$, the thickness of the plate films.  If small, it is easy to reach in simulation. 

Finally, the Thomas-Fermi screening length $\Lambda \sim v_F/\omega_p$  is much smaller, on the order of Angstrom.   In
numerical sum over ${\bf p}_\perp$, we take typically $L=60$, with integration bandwidth $|E| < 7t$ and 400 points to sample
the energy. 
 
\subsection{Property of $\Pi$ near surface}

\begin{figure} 
 \centering
  \includegraphics[width=0.8\columnwidth]{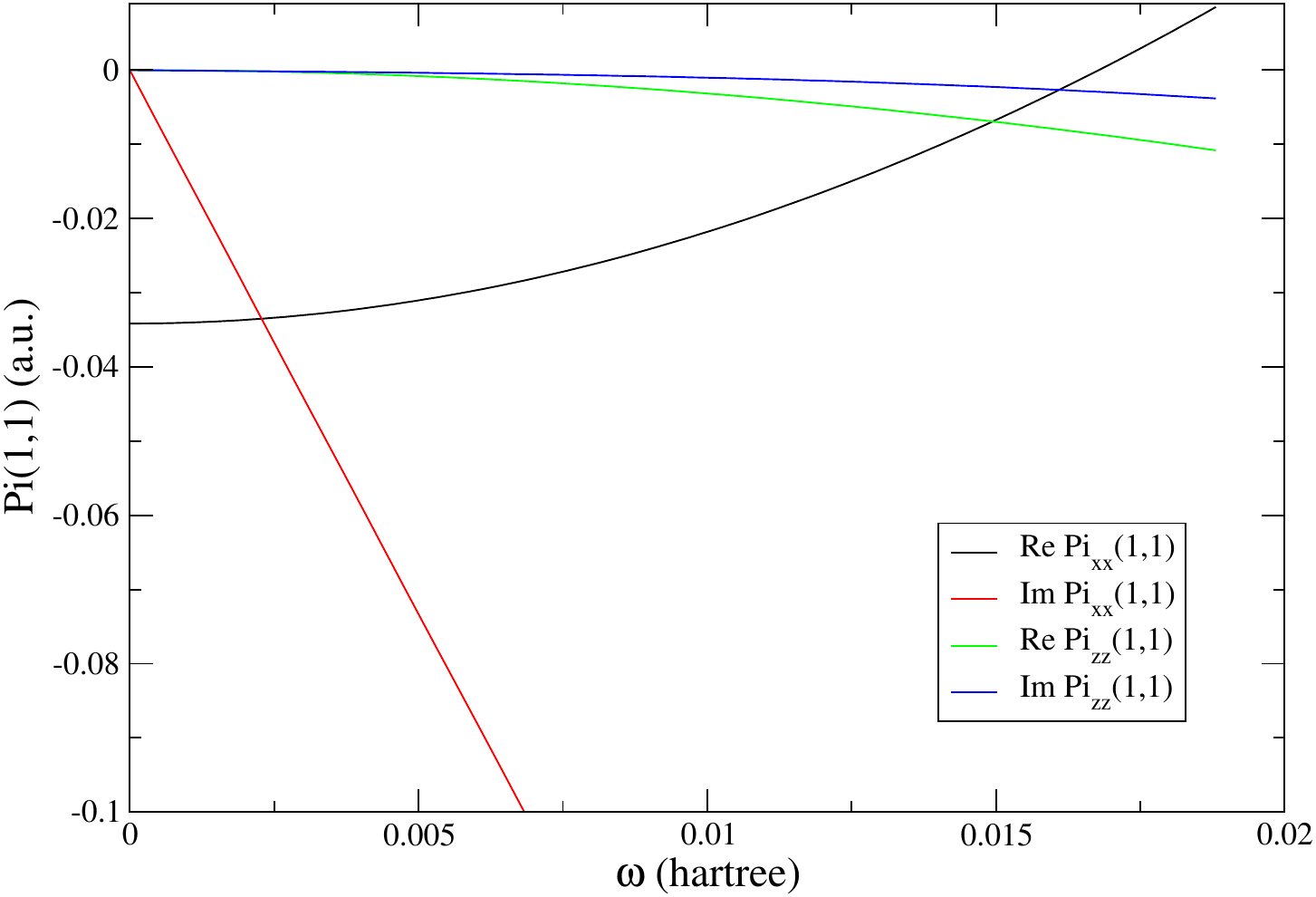}
  \caption{\label{fig-Piz11-w}The polarizability on the surface, $\Pi_{\alpha\alpha}(\omega,{\bf q}_\perp = {\bf 0},1,1)$.  It is
  not zero at zero frequency in the transverse directions.}
\label{coulomb}
\end{figure} 

The gauge invariance requires a constant vector potential $\bf A$ to produce zero induced current.
In a real space representation of $\Pi$, the gauge invariance implies $\int \Pi(\omega\!=\!0,{\bf r}, {\bf r'}) d^3 {\bf r}' = 0$. 
For translationally invariant systems, this is $\Pi=0$ when $\omega = 0$ and ${\bf q}={\bf 0}$, as in
Eq.~(\ref{eq-gauge-0}).
The RPA expressions $\Pi^{ {\rm RPA}}$ and the diamagnetic expressions $\Pi^{\rm dia}$ above are based on
the lowest order of perturbation theory and not an exact result.  For $\Pi_{zz}$, $\Pi_{xy}$ or $\Pi_{zy}$ the cancelation still probably is exact,
in that numerically $\Pi = \Pi^{\rm PRA} + \Pi^{\rm dia}$ is 0 for all $(j,j')$ to machine precision ($10^{-16}$ in atomic units).
This is mysterious, and the authors do not know the deep reason.   In transverse direction, however, $\Pi_{xx} = \Pi_{yy}$ is not zero. 
The matrix is approximately tridiagonal with $\pi_0 \approx 0.035\,$a.u.~on the first off-diagonals when $|j-j'|=1$, and the diagonals are determined by the gauge invariance sum rule
$\sum_{j'} \Pi(j,j')=0$.     Thus, the matrix in $(j,j')$ at 
$\omega = 0$ and ${\bf q}_\perp = 0$ takes the form
\begin{equation}
\label{eq-p0-matrix}
\left[ \begin{array}{rrrr}
-\pi_0 & \pi_0 & 0 & \cdots \\
\pi_0 & -2\pi_0 & \pi_0 & 0 \\
0 & \pi_0 & -2\pi_0 & \pi_0 \\
\vdots & 0 & \pi_0 & -2\pi_0 
\end{array}\right].
\end{equation}
The diagonal values at $j=j'=1$ are plotted in Fig.~\ref{fig-Piz11-w}.  Away from ${\bf q}_\perp = 0$, the magnitude of $\Pi_{xx}$ decreases
and $\Pi_{zz}$ increases smoothly.  But the values remain small.

If this incomplete paramagnetic and diamagnetic cancelation is not a random phase approximation artifact and is an actual physical effect (look at Eq.~(\ref{eq-gauge-0}),
where the cancelation occurs only at ${\bf q} ={\bf 0}$; the residue diamagnetic term represents a magnetic response), then this numerical result has a consequence: it might be
the point of resolution of the Drude vs.\ plasma model controversy.   
The high-temperature limit of the attractive Casimir pressure (the Matsubara frequency $\omega_n=0$ contribution) is given by
\cite{Lifshitz_1956,Klimchitskaya09}
\begin{equation}
\label{eq-thermal-Casimir}
\frac{P}{k_BT}  = \int_0^{\infty} \frac{dq}{2\pi} q^2 \sum_{\sigma=s,p} \frac{r_\sigma^2(0,q) e^{-2dq}}{1 - r_\sigma^2(0,q)e^{-2qd}}. 
\end{equation}
We have $r_s = 0$ and $r_p = -1$ for the Drude model.   On the other hand, if $\Pi_{xx}$, $\Pi_{yy}$ is not zero while the sum over $j'$ equal zero
to be compatible with gauge invariance,
the polarizability can be approximated in the continuum limit as
\begin{equation}
\Pi_{zz} = 0,\quad \Pi_{xx} = \Pi_{yy} = \frac{b}{\mu_0} \frac{d^2\ }{dz^2}.
\end{equation}
They are differential operators (except that the boundary condition at the surface is not reflected).
Here $b =\mu_0 \pi_0/a \approx  4.8\times 10^{-6}$, $a$ is lattice constant, and $b$ is  dimensionless. 
Focusing only on the $s$-polarization since $r_p$ is already $-1$, working at $\omega \to 0$, we find the reflection
coefficient is not zero, but similar to the plasma model case (but not $q$ dependent), as, by solving the scattering 
problem for the wave equation, $\epsilon_0(\omega^2 - c^2 q^2 + c^2 d^2/dz^2) {\bf A} = \Pi {\bf A}$, which is equivalent to a renormalized speed of light, $c^2 \to (1 - b)c^2$, 
\begin{equation}
r_s = \frac{\sqrt{1- b} - 1}{\sqrt{1- b} + 1} \approx -\frac{b}{4} = -1.2 \times 10^{-6}.
\end{equation}
Unfortunately, this is too small to have a real effect.  

\subsection{Diamagnetic materials}

But the above result is not quite correct because the gauge sum rule and the boundary condition precisely at the interface is
not taken care of.  The reflection critically depends on the boundary condition.  Now, we do it in a discrete space and consider it more carefully.   First, let's consider the perfect diamagnet,
i.e., $\Pi = \Pi^{\rm dia}$ is a constant.   Then, the equation for $s$-polarization should be
\begin{equation}
\epsilon_0 \left( \omega^2 - c^2 q_\perp^2 + c^2 \frac{d^2\ }{dz^2} \right) A = \Pi^{\rm dia} A.
\end{equation}
Here $A$ is either $A_x$ or $A_y$. 
The waves in the diamagnet are strongly damped with a dispersion relation $(\omega/c)^2 = (k_z')^2 + q_\perp^2 + \mu_0 \Pi^{\rm dia}$.
The diamagnetic term introduces a decay scale, which is the London screening length, $\delta_L = (\mu_0 \Pi^{\rm dia})^{-1/2}$ about 
100\,a.u.\ for our parameters.   The reflection coefficient is 
\begin{equation}
r_s = \frac{k_z - k_z'}{k_z + k_z'} \approx -1
\end{equation}
in the $\omega \to 0$ and $q_\perp \to 0$ 
limit, because the $\Pi^{\rm dia}$ term protects $k_z'$ from going to 0.  This is the plasma model result, while the Drude model gives $r_s = 0$. 

But our system is not this.  Our $\Pi$ is nonlocal, so in a continuum formulation, we can write the equation as
\begin{equation}
\epsilon_0 \left( \omega^2 - c^2 q_\perp^2 + c^2 \frac{d^2\ }{dz^2} \right) A(z) = \int dz' \Pi(z,z')  A(z').
\end{equation}
Let's discretize this equation on both sides, and the second derivative with respect to $z$ is represented by the
 central difference formula,  $( A_{j-1} - 2A_j + A_{j+1})/a^2$, here $z=aj$, for all $j$.   We assume the metal 
occupies the sites $j=1,2,\cdots$, and $j=0,-1, \cdots,$ are vacuum.   Then, the right-hand side takes the form of
matrix multiplication by (\ref{eq-p0-matrix}), but only for the subblock for $j\ge 1$.   Thus, we can write a set of
linear equations, as
\begin{align}
&\qquad\ \  A_{j-1} + (-2 + \delta) A_j + A_{j+1}  = 0, \quad j \leq  0, \nonumber \\
&A_0 + (-2+\delta + b) A_1 +(1-b) A_2= 0, \quad j = 1, \label{eq-j1}  \\
&(1\!-\!b) A_{j-1} + (-2\!+\!\delta\! +\!2 b) A_j + (1\!-\!b) A_{j+1} = 0, \quad j> 1. \nonumber 
\end{align}
Here we have defined $\delta = k_z^2 a^2$, $k_z^2 = (\omega/c)^2 - q_\perp^2$, and $b = \mu_0 \pi_0/ a$.

We seek for the scattering solution with the ansatz:
\begin{equation}
A_j = \left\{  \begin{array}{ll}
\zeta^j + r_s \zeta^{-j}, & \mbox{if $j<1$} \\
t_s \,\zeta'^{(j-2)}, &\mbox{otherwise.}
\end{array}
\right. 
\end{equation}
Here $r_s$ is the reflection coefficient, and $t_s$ is the transmission coefficient in $s$ polarization.  For the propagating modes,
$|\zeta = |\zeta'| = 1$. 
In vacuum, for $j<1$, the wave is determined by the equation with $\zeta$ satisfying 
\begin{equation}
\zeta + \bigl(-2 + \delta\bigr) + \frac{1}{\zeta} = 0,
\end{equation}
while for the metal when $j>1$, it is determined by a similar equation with $\delta$ replaced by
$\delta' = \delta/(1-b)$, producing the solution $\zeta'$.   A non-zero $b$ signifies that the metal is different from
the vacuum.  [All this discussion is assumed near $\omega = 0$]. 

Critically, exactly at the boundary, $j=1$, the equation is different, given by (\ref{eq-j1}).
The ansatz can be verified to be true, and the set of linear equations can be solved to give
$A_0 = 1 + r_s$, $A_1 = t_s/\zeta' = \zeta+ r_s/\zeta$.  The last equality is obtained 
from the equation when $j=0$.  Finally, the reflection coefficient is
\begin{align}
r_s & = - \frac{1+u \zeta}{1+u/\zeta}, \\
u &= -2 +\delta + b + (1-b) \zeta'.
\end{align}
We can take the continuum limit with this formula.   In the continuum limit, $\zeta = e^{i k_z a} \to 1$,
and $\delta = k_z^2 a^2 \to 0$, and $b = \mu_0 \pi_0/a \sim a^2$, (this is because
$\pi_0 \sim a^3$, the amount of $\Pi$ per lattice site).  As a result, we find
\begin{equation}
r_s \approx \frac{1}{4} b, \qquad a \to 0,
\end{equation}
i.e., it is still fairly small.  So, the idea of a finite $\Pi_{xx}$ producing a perfect diamagnet doesn't work.  The main reason is that the magnetic response proportional to $\mu_0 \sim 10^{-3}\,$a.u.\ is too small \cite{Esquivel04,Henkel2411}. 

We see that a perfect diamagnet of $\Pi^{\rm dia} \approx {\rm const}$ gives $r_s=-1$, but $\Pi \propto d^2/dz^2$ gives 0.
We might have an intermediate case of 
\begin{equation}
\label{eq-Klim}
\Pi(\omega \to 0, {\bf q}) \approx \frac{e^2 n \tau}{m} v^T q_\perp. 
\end{equation}
Here, $\tau$ is the Drude relaxation time, and $v^T$ is a fitting parameter with the dimension of velocity.   This is the form given by 
Klimchitskaya and Mostepanenko \cite{Klimchitskaya20,Klimchitskaya22PRA}.   Indeed, if we use this form of $\Pi$ as valid in the bulk, we obtain $r_s = -1$ again, reproducing
the plasma result,  since now $k_z' = \sqrt{ - q_\perp^2 -  q_\perp v^T \mu_0\, \sigma} \propto \sqrt{-q_\perp}$ 
goes to zero slower than $q_\perp$. 
The question is if the form is physically reasonable. 

\begin{figure} 
 \centering
  \includegraphics[width=0.8\columnwidth]{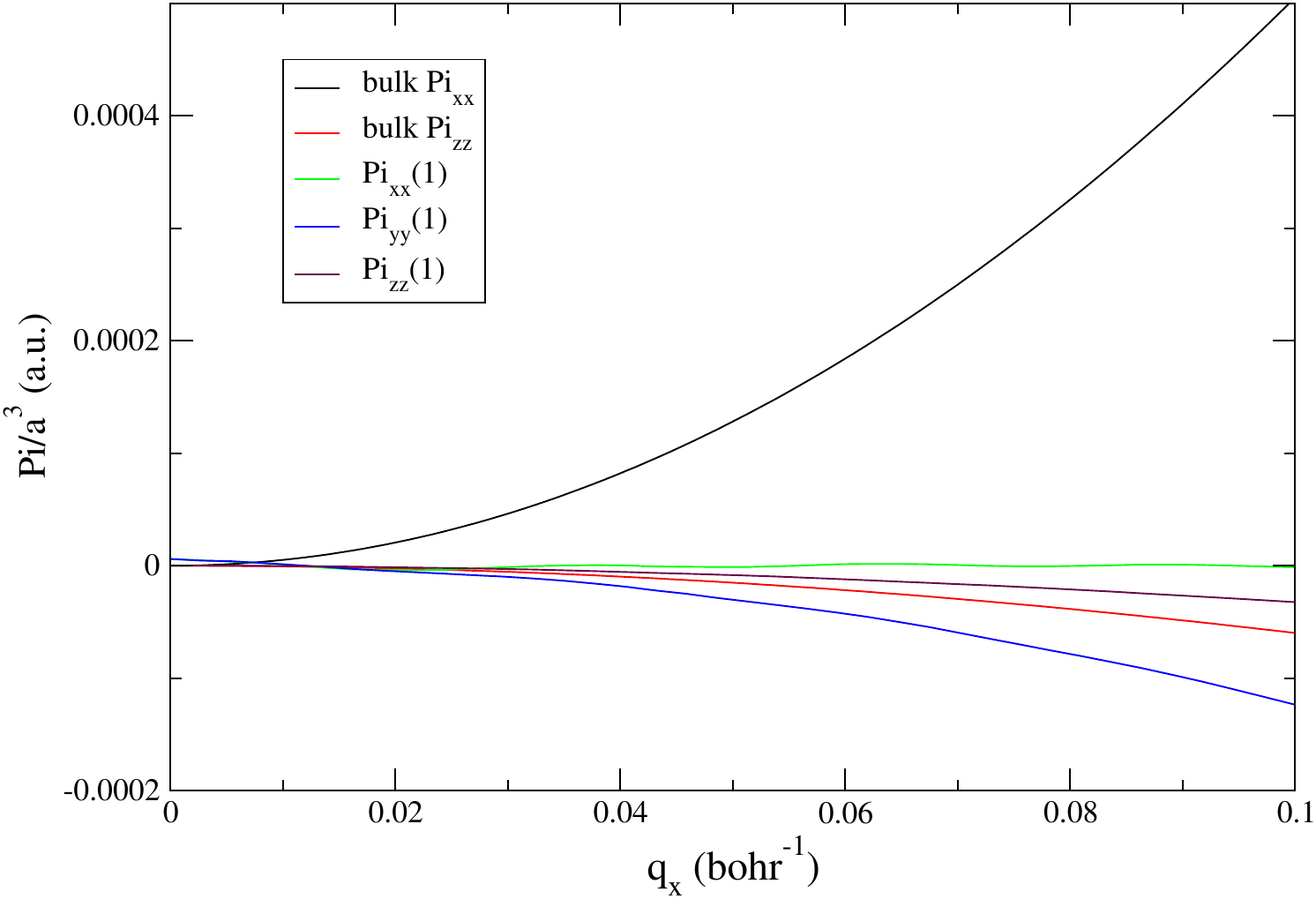}
  \caption{\label{fig-Pixx-zz-q}The polarizability at zero frequency on the surface and in bulk at $q_y=0$, as a function of 
  $q_x$.   In this geometry, transverse means $y$ and $z$ direction.  The curves can be fitted to $q_x^2$.}
\label{coulomb}
\end{figure} 

We check the possible $q_\perp$ dependence numerically and present in Fig.~\ref{fig-Pixx-zz-q}.  We compute $\Pi$ at $\omega=0$ and
$q_y = 0$.
We use Eq.~(\ref{eq-RPA-Pi}) at $q_y=q_z = 0$ for the bulk result.  For the surface, we compute $\sum_{j'} \Pi(1,j')$ at the first 
layer.  The values are small compared to $e^2 n/m = 0.0087\,$a.u.. 
According to Eq.~(\ref{eq-Klim}) with $v^T$ of order Fermi velocity, we should see a slope of order 1; the actually
slope is $\sim 10^{-4}$. 
 Based on the numerical data, it does not seem to support a linear
$q_\perp$ diamagnetic response in our system.

\subsection{When do we approach the local Drude model?}
\begin{figure} 
 \centering
  \includegraphics[width=0.525\columnwidth]{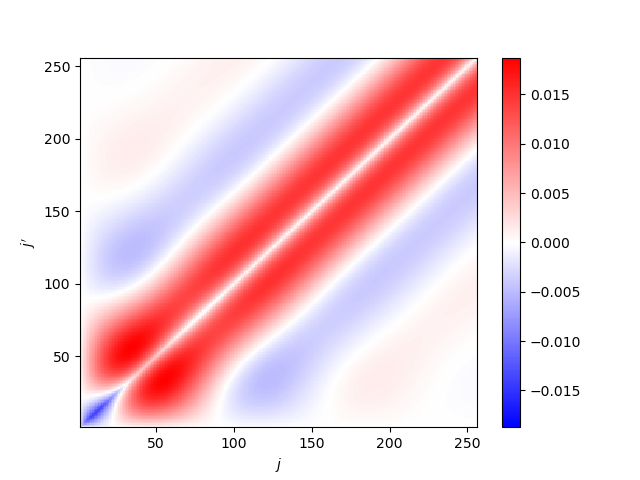}\includegraphics[width=0.525\columnwidth]{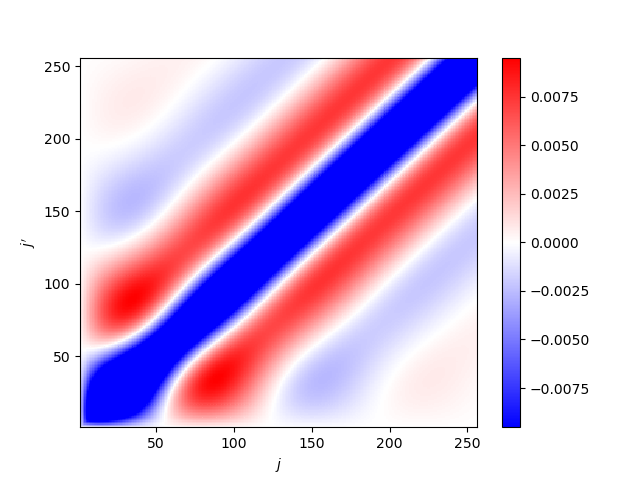}
  \caption{\label{fig-Pizz-6kT}The polarizability $\Pi_{zz}(\omega = 6k_BT,{\bf q}_\perp = {\bf 0}, j,j')$ in the distance variables $(j,j')$.  Left is the real part; right is the imaginary part.}
\label{coulomb}
\end{figure}

We check how well the $\Pi$ expression recovers the Drude model as we go away from the surface.  Away from
$\omega=0$, $\Pi$ displays non-local behavior in $z$.   Figure~\ref{fig-Pizz-6kT} displays a density plot for the real
and imaginary part of $\Pi_{zz}$ in the index $(j,j')$ in $z$ direction at $\hbar \omega  = 6k_B T$ ($T = 300\,$K) 
and transverse wavevector ${\bf q}_\perp = {\bf 0}$.   They are not localized at diagonal, indicating it is not of the form $\delta(z-z')$. 
The scale of $\Pi$ in $z$ is on the order of 100 lattice spacings, decaying to 0 rather slowly at lager $|j-j'|$. 

\begin{figure} 
 \centering
  \includegraphics[width=0.8\columnwidth]{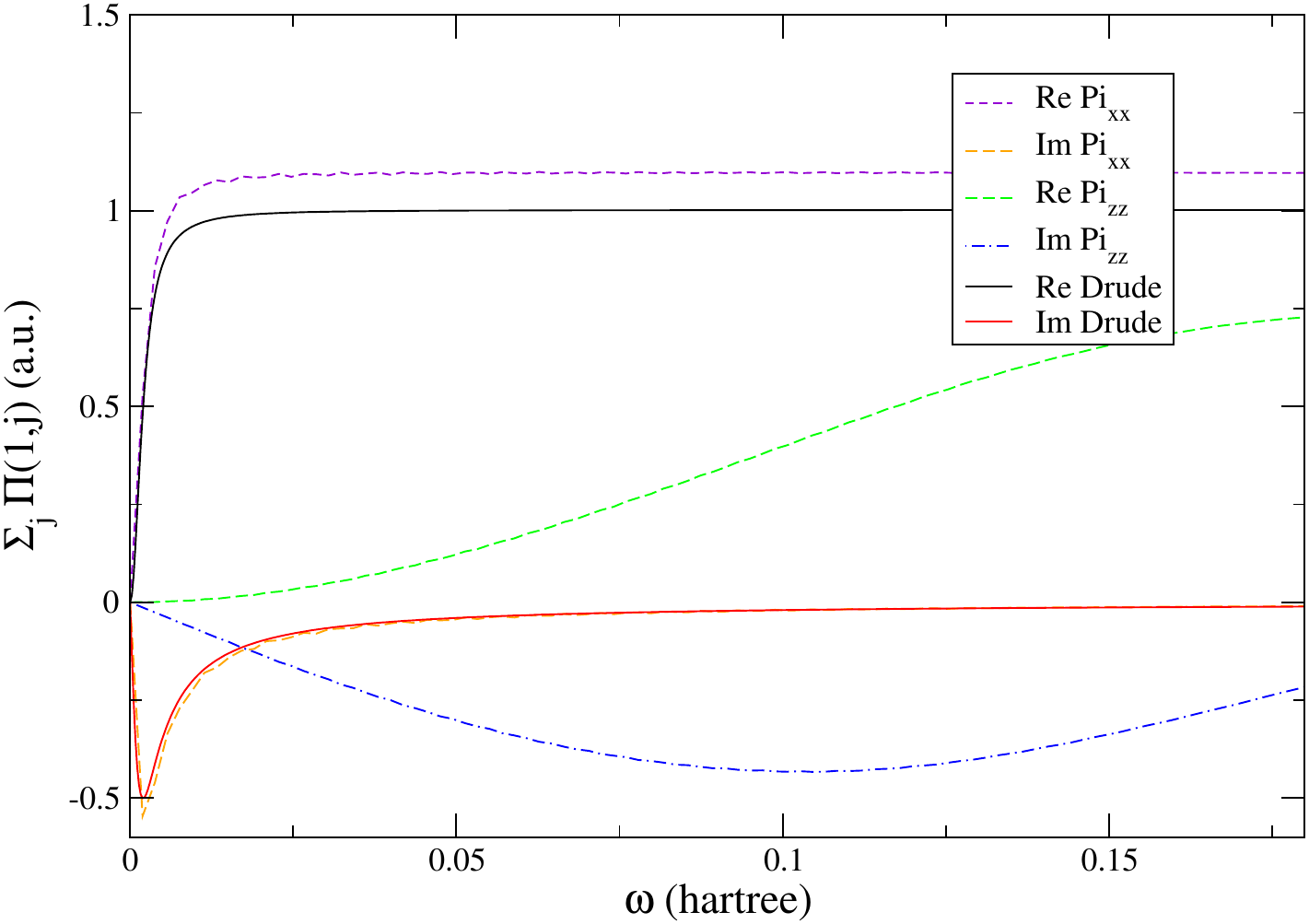}
  \caption{\label{fig-Piz-sum-z1}The column sum of polarizability, $\sum_{j'}\Pi_{\alpha\alpha}(\omega,{\bf q}_\perp={\bf 0}, j=1,j')$, as a function of the frequency
$\omega$ in atomic units (hartree).  The dotted and dashed lines are for the real and imaginary parts of the surface model.  The solid lines are 
from the local Drude model, Eq.~(\ref{eq-local-Drude-Pi}).}
\label{coulomb}
\end{figure} 

In Fig.~\ref{fig-Piz-sum-z1}, we present the column sum of $\Pi$ as a function of the frequencies.  The sum over $j'$ has the 
effect of computing $\Pi$ also in longwave in $z$ direction, $q_z=0$, while $j$ indicates how far we are from the surface.   The figure gives the
result in the surface layer.  In the transverse
directions, $\Pi_{xx}=\Pi_{yy}$, the sum is well described by the Drude model of Eq.~(\ref{eq-local-Drude-Pi}).  But in 
the $z$ direction, it differs hugely from the Drude model.  Any good analytic theory of the surface $\Pi$ has to deal with
this correctly. 

We can introduce two relaxation times to explain these results.    For the transverse directions, $x$ or $y$ components here
since ${\bf q}_\perp \approx {\bf 0}$, the Drude model works fine with the bulk relaxation time $\tau_T = \hbar/(2\eta)$,  
unaffected by the presence of the surface since this is just like a 2D metal.   Charge conservation is unrelated to them,
see Eq.(\ref{eq-PtoPi}) below.  The longitudinal component ($z$ direction) is greatly modified with a much shorter
relaxation time of order $\tau_L \approx \hbar / t$, about 40 times smaller.  

\begin{figure} 
 \centering
  \includegraphics[width=0.8\columnwidth]{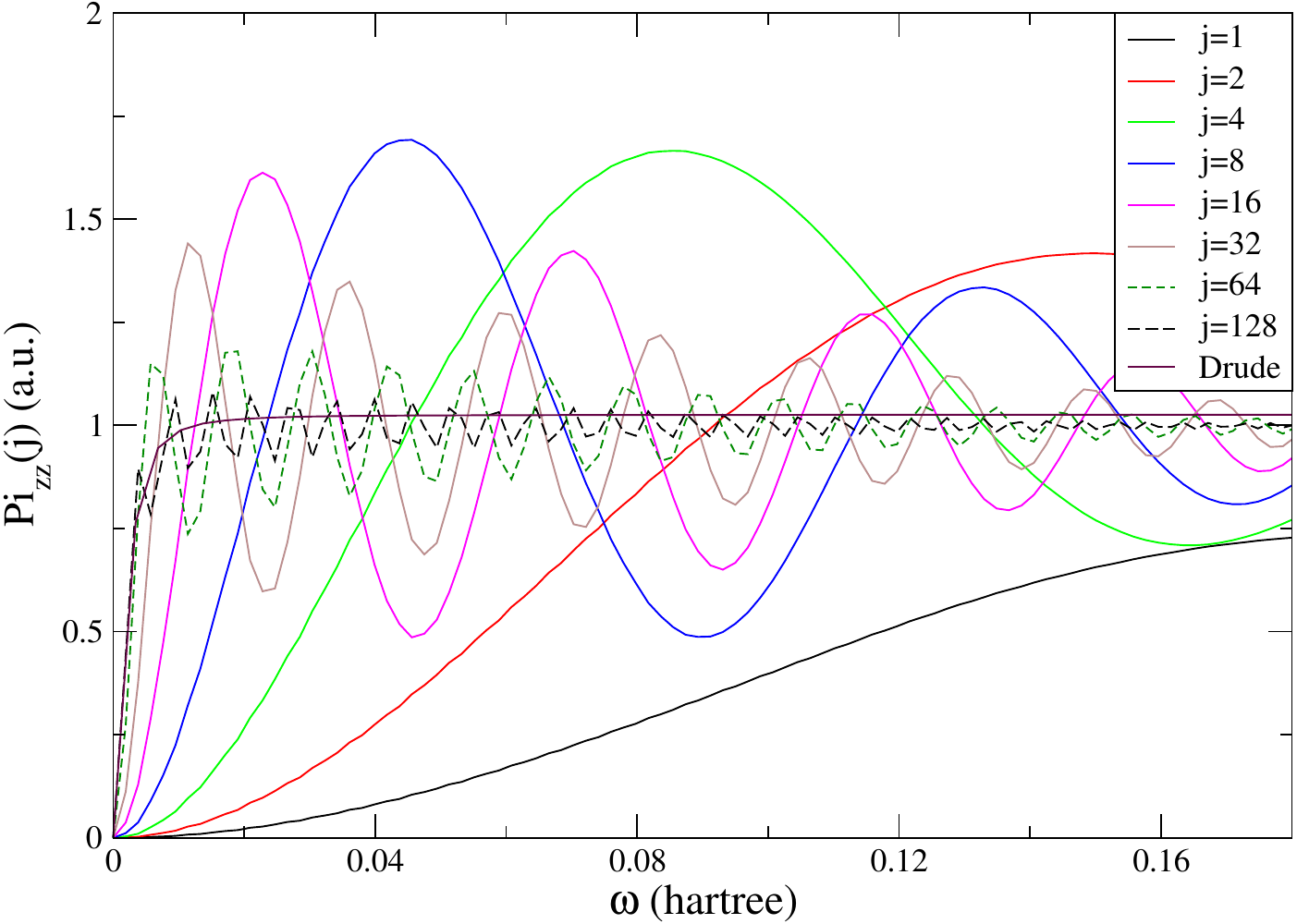}
  \caption{\label{fig-Pizz-j}The sum of polarizability $\sum_{j'}\Pi_{\alpha\alpha}(\omega,{\bf q}_\perp={\bf 0}, j,j')$ as a function of the frequency
$\omega$ for variance distance $j$ away from the surface.}
\end{figure}

Due to the oscillations away from the first layer (see Fig.~\ref{fig-Pizz-j}), the marked decrease for $\Pi_{zz}$ at small frequencies can not simply be explained by a small $\tau$.   It is better to work with $P$ to be discussed in the next section. 
Intuitively, these oscillations come from the surfaces Green's functions.   The solution of the surface Green's function
for each fixed $j'$ comes from traveling wave away from $j'$ to infinity for the index $j>j'$.  But between $1 \leq j < j'$, it is an
interference form of $A \lambda^j + B \lambda^{-j}$.  The constants $A$ and $B$ depend on $j'$.   However, $\Pi$ is
a convolution of two surface Green's functions in energy $E$.  So, the relationship between $g$ and $\Pi$ is complicated.  

In Fig.~\ref{fig-Pizz-j}, we have a busy plot for the real part of $\sum_{j'} \Pi_{zz}$ for various distances $j$ away from the surface.
It approaches the Drude result in an oscillatory fashion.  Typically, convergence is reached in about 100 lattice spacings.   
The imaginary part is similar, so we will not show it.  It is a curious observation that the period of oscillation in frequency 
decreases by a factor of 2 whenever the distance away from the surface increases by a factor of 2.  The oscillation amplitudes become smaller as we go away from the surface. 

\section{Charge-charge correlation}
For the pure longitudinal or Coulomb interaction, the associated $\Pi$ is the scalar field version
for charge correlation which we denote by $P$. 
We motivate here the fitting form 
\begin{equation}
P(\omega, {\bf q}_\perp, j,j') \approx  (a + i b \omega)\delta_{jj'} 
\end{equation}
for the next section based on the periodic system.  This form is inspired by the numerical data we obtained
from the full calculation of $P$ using the surface Green's function.  In terms of $(j,j')$, the values of $P$ are nearly diagonal
with off-diagonals about 10 times smaller at $\hbar\omega = k_kT$.  It weakly depends on ${\bf q}_\perp$.  
In fact, this form can be argued by symmetry $P(\omega)^* = P(-\omega)$, plus analyticity. 
The RPA expression in mode space is (in a lattice periodic system) \cite{Giuliani-Vignale05}
\begin{align}
P(\omega, {\bf q}) &= 
\frac{2e^2}{N} \sum_{{\bf k}}  \frac{ f(\epsilon_{\bf k} + i \eta) - f(\epsilon_{{\bf k}+{\bf q}} - i\eta)
}{ \hbar \omega + 2 i \eta + \epsilon_{\bf k} - \epsilon_{{\bf k} + {\bf q}} }.
\end{align}
With a parabolic dispersion relation, this gives the Lindhard function (when $T\to 0$ and $\eta \to 0$).    The 
Lindhard function is singular near ${\bf q}=0$ and $\omega=0$.  For large $\omega$, $\bf q$ small, we can use
Taylor expansion to show $P(\omega, {\bf q}) \sim q^2/\omega^2$.   But this limit is inappropriate
for near-field heat transfer as it is dominated by small $\omega$ ($< \eta/\hbar$).  If we use ${\bf q}={\bf 0}$ but for
finite $\omega$, we find
\begin{align}
P(\omega, {\bf 0}) & \approx \frac{2 e^2}{N} \sum_{\bf k} \frac{2 i \eta \, df/d\epsilon_{\bf k}}{\hbar \omega +2 i\eta} \nonumber \\
& = -e^2 {\rm DOS}(\mu) \frac{ (2\eta)^2 + 2 i \eta \hbar \omega}{(\hbar \omega)^2 + (2\eta)^2},
\end{align}
where ${\rm DOS}(\mu)$ is the density of states per site at the Fermi level. 
This gives a constant for the real part and is proportional to $\omega/\eta$ for the imaginary apart, which
is a major contribution for 2D metal plates \cite{Xie2412arxiv}.  Note that this limit cannot be viewed as a Drude model. 

\section{Heat transfer in the non-retardation limit}
In this section, we consider the heat transfer in the limit of speed of light $c \to \infty$.   In this limit, it is equivalent to
consider only the Coulomb interaction and the scalar potential.  The Polder-van~Hove result for the power density reduces to \cite{Joulain05}
\begin{equation}
\label{eq-Landauer}
I = \int_0^\infty \frac{d\omega}{2\pi} \hbar \omega\,( N_1 - N_2) \int \frac{d^2 {\bf q}_\perp}{(2\pi)^2} T({\bf q}_\perp, \omega), 
\end{equation}
where $N_1$ and $N_2$ are the Bose function at temperature $T_1$ and $T_2$ of the two blocks, the transmission coefficient is
\cite{Zhang18prb,wang_transport_2023}
\begin{equation}
\label{eq-c-inf-T}
T({\bf q}_\perp, \omega) = \frac{ 4\, {\rm Im}\, r_1\, {\rm Im}\, r_2\, e^{-2qd}}{|1 - r_1 r_2 e^{-2qd}|^2} ,
\end{equation}
where $r_1$ and $r_2$ $(= r_p {\rm\ or\ }  r_c)$ are the reflection coefficients, $d$ is the distance between two blocks or plates, $q = |{\bf q}_\perp|$.
Only the $p$-polarization contributes, and the $s$-polarization is zero within this limit. 
For a semi-infinite block of metal at $c\to \infty$,  $r_p = (k'_z-\epsilon k_z)/(k'_z+\epsilon k_z) \approx (1-\epsilon)/(1+\epsilon)$   if we use the Drude model local dielectric function
$\epsilon$  \cite{Ford84}, 
it is independent of the wavevector $\bf q$; the result trivially gives a $1/d^2$ distance dependence, which is well-known \cite{Volokitin07,Chapuis08,Ottens11}.  

Numerical simulations using the cubic lattices show that this $1/d^2$ Drude model result is not true. 
To explain this discrepancy,  
we realize that the metal plates have $P = - e^2 D_0 (1 + i \tau \omega)$ (per site for a lattice model, taking the continuum limit
means we need to replace $P$ by $P/a^3$ to be per volume).  Here $\tau$ is an effective relaxation time much smaller than
the bulk value $\hbar/(2\eta)$ ($12\,$a.u.~vs.\ $500\,$a.u.\ for our model parameters). 
 $D_0$ is the electron surface DOS per site. 
To work out the scattering problem and find the reflection coefficient in this setting (we call $r_c$),  we have to solve
the Poisson equation with an induced charge,  $P \phi$,  in the spirit of Ref.~\cite{Henkel2411}. 
In this new setting, i.e., under the scalar potential $\phi$ in the small $q$ limit valid for long distance, we find (see Sec.~\ref{secIX} for a derivation)
\begin{equation}
r_c  = \frac{q-k}{q+k}  \approx  -1 +2 \Lambda q - i \Lambda \tau \omega q, 
\label{r-scalar}
\end{equation} 
where $k = \sqrt{ q^2 - P/(\epsilon_0 a^3)}$, and we have defined the Thomas-Fermi screening length by
\begin{equation}
\Lambda = \sqrt{\frac{a^3 \epsilon_0}{e^2 D_0}}.
\end{equation}
Using the expression for the reflection coefficient $r_c$ for Eq.~(\ref{eq-c-inf-T}), setting the denominator to 1 in the transmission function in Eq.~(\ref{eq-c-inf-T}) valid for long distance, we can perform the
integrals, and obtain
\begin{equation}
\label{eq-I-IBB}
\frac{I}{I_{\rm BB}} = \frac{3}{2} \cdot \frac{\bigl(\Lambda \tau c\bigr)^2 }{d^4}.
\end{equation}
Here $I_{\rm BB}$ is the black-body result when $T({\bf q}_\perp, \omega)=2\, \theta(\omega -cq)$, here $\theta$ is the unit step function.

The above large $d$ asymptotic behavior is not captured in the Drude model, but it is a natural consequence of the small $q$ behavior for $P(\omega, {\bf q})$.  The Drude model, valid in large $\omega$, is given by
\begin{equation}
P(\omega, {\bf q}) \approx q^2 \frac{ e^2 n}{m} \frac{1}{ \omega( \omega + 2 i \eta/\hbar)},
\end{equation}
which vanishes in the long-wave limit, but the plate result does not vanish in this limit and is given by the negative of the electron density of
states.   The scalar $P$ and the vector field version of $\Pi$ are related by  $\omega^2 P = {\bf q} \cdot \Pi \cdot {\bf q}$ due to charge
conservation.  This relation is valid in the bulk but becomes singular near the surface, 
\begin{equation}
\label{eq-PtoPi}
\omega^2 P({\bf q}_\perp, z,z') = {\bf q}_\perp \cdot \Pi \cdot {\bf q}_\perp + 
\frac{\partial \ }{\partial z} \frac{\partial \ }{\partial z'} \Pi_{zz}({\bf q}, z,z').
\end{equation}
We see that in order to have the scalar polarization $P \sim a + i b \omega$, the imaginary part of $z$ component must increase 
cubically with frequency 
as ${\rm Im}\,\Pi_{zz} \sim \omega^3$, instead of the Drude model of linear $\omega$. 

\section{Numerical result of cubic lattice blocks}
We verify the $1/d^4$ dependence numerically with semi-infinite cubic lattice blocks.  The cubic lattice has a lattice constant $a$ with periodic
boundary conditions in the $(x,y)$ plane,  and open boundaries at the ends.  The two blocks have a gap of $d$.   Due to periodicity in the transverse direction, we work in the Fourier space ${\bf q}_\perp$ in the plane and in real space in the $z$ direction. The Landauer formula, Eq.~(\ref{eq-Landauer}), still gives the heat current with the transmission coefficient given by the Caroli formula \cite{jiang17}:
\begin{equation}
T({\bf q}_\perp, \omega) = {\rm Tr} ( D^r \Gamma_2 D^a \Gamma_1),
\end{equation}
where the trace is over the layer label $z=a j$.   The spectrum function is $\Gamma_\alpha = i(P_\alpha - P^\dagger_\alpha)$, $\alpha = 1,2$ label
the left and right side of the gap.    $D^r = v + v (P_1 + P_2) D^r$ is the scalar field retarded Green's function; it is a matrix equation in discrete $z$ space.
The advanced Green's function is $D^a = (D^r)^\dagger$.  The free scalar-photon Green's function associated with the Poisson equation is 
\begin{equation}
\label{eq-coulomb-kernel}
v({\bf q}_\perp, \omega,z,z') = \frac{e^{-q |z-z'|}}{2 \epsilon_0 q\, a^2},\quad q=|{\bf q}_\perp|.
\end{equation}
The extra factor of $a^2$ in the denominator is because our definition of $P$ is per site.  We note $vP$ or $D^r\Gamma_\alpha$ is dimensionless. 

\begin{figure} 
 \centering
  \includegraphics[width=1.0\columnwidth]{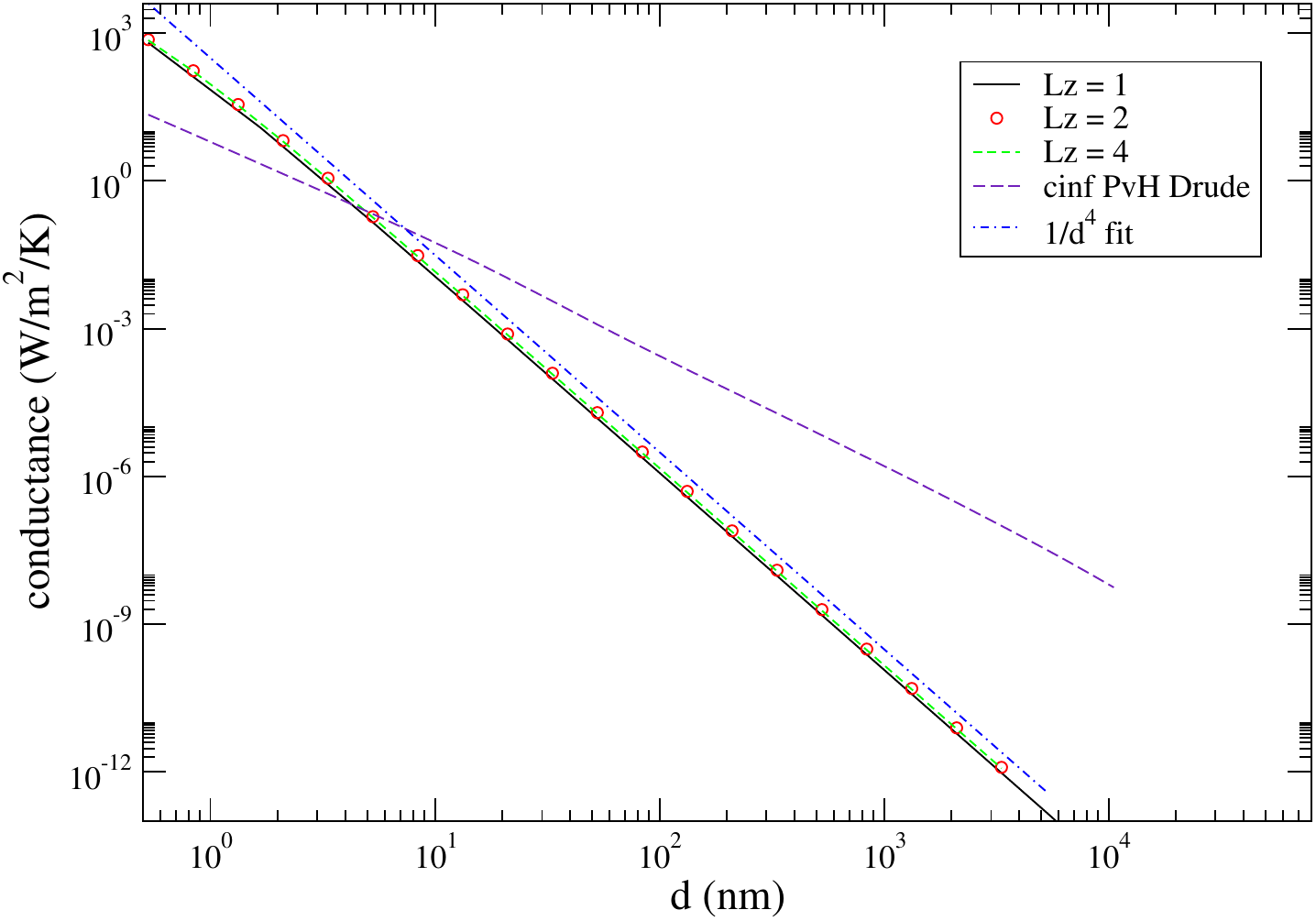}
  \caption{The heat conductance at 300$\,$K as a function of the gap distance $d$.  $L_z$ labels the thickness of the blocks
treated explicitly for Coulomb interaction.  The dashed line labeled ``cinf PvH Drude'' is computed from the Drude model with the PvH theory at $c\to \infty$. The parameters $D_0 = 5.58\,$a.u. and $\tau = 12.0\,$a.u.\ are obtained by
fitting $P$ at the first layer; the resulting prediction is labeled as $1/d^4$ fit. }
\label{coulomb}
\end{figure} 

In Fig.~\ref{coulomb}, we present the conductance, which is defined as
\begin{equation}
G = \frac{dI}{dT} = \lim_{\Delta T \to 0} \frac{I(T_1 = T + \Delta T, T_2=T)}{\Delta T}.
\end{equation}
This is obtained by replacing $N_1 - N_2$ by $dN/dT$.   We see that the results converge quickly with the layer thickness $L_z$.  
 The result for $L_z = 4$ and 8 (not plotted) is nearly indistinguishable.  This is due to the smallness of the Thomas-Fermi screening length.  
 It agrees very well with the $1/d^4$ dependence starting from the nm range. 
However, the analytic formula, Eq.~(\ref{eq-I-IBB}), predicts a value about 3 times larger (perhaps caused by the assumption that
$P$ is strictly local).
The heat power is quite low; at $0.53\,$nm the 
value is $716.9\,$W/(m$^2$K).  This is to be compared with the full electromagnetic contribution from 
PvH theory of Drude model of Au of  $3930.4\,$W/(m$^2$K) at the same distance.   The Drude model non-retardation
limit  (i.e., at $c\to \infty$) is even smaller,  $22.1\,$W/(m$^2$K), see Fig.~\ref{fig-drude-compare}.
This contradicts the belief \cite{Mahan17} that Coulomb interactions are most important 
at short-length scales.   This is not true for intensely screened systems like metal. 

Although the effect of $1/d^4$ is small for Au, we expect it will be larger for poor conductors with smaller carrier density,
for example, a heavily doped Si. If we decrease $D_0$ by a factor of 2, the current $I$ will increase by a factor of 2.  
So it is still observable.   

\section{Comparing PvH theory with NEGF solution method}
\begin{figure} 
 \centering
  \includegraphics[width=1.0\columnwidth]{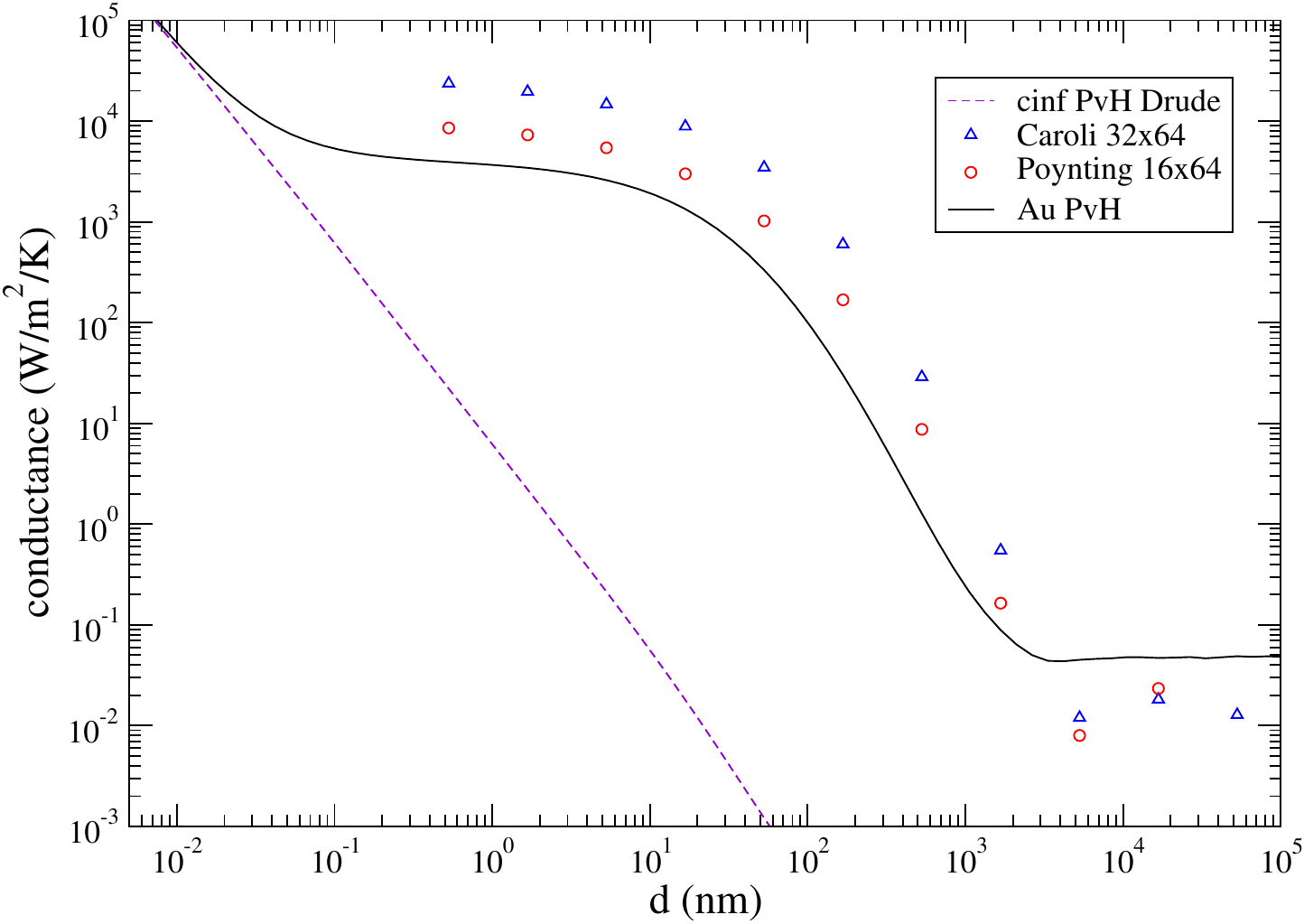}
  \caption{\label{fig-drude-compare}Compare the PvH theory (solid black) with the NEGF results for the local Drude model.  The parameters are for gold with $\hbar \omega_p = 9.02\,$eV, $\hbar/\tau = 54\,$meV.  The thickness of the blocks: the red circles  $L_z=16\times 64$, 
blue triangles  $L_z = 32\times 64$,  using the transverse free Green's function. Thus, the longitudinal Coulomb contribution is excluded.  To be able to reach large sizes, we scaled
the lattice constant $a$.   The first number is the scaling factor; the second number is the actual grid points.  The dashed line
labeled ``cinf PvH Drude'' is the same as in Fig.~\ref{coulomb}.}
\label{config}
\end{figure}

In PvH theory, we calculate the reflection coefficients for one interface and then put the reflection coefficients into the Landauer formula.  In the nonequilibrium Green's function 
(NEGF) method \cite{zhang_microscopic_2022,wang_transport_2023,Zhu24}, the Landauer formula is the same, but the transmission function
${\rm Tr} ( D \Gamma_2 D^\dagger \Gamma_1)$ is calculated numerically by solving the Dyson equation,
$D = v + v \Pi D$.   For a vector theory, $D$ is a 3 by 3 matrix in the direction indices (dyadic Green's function). 
Alternatively, we can also evaluate the average Poynting vector.   If we 
use the local Drude model $\Pi$, Eq.~(\ref{eq-local-Drude-Pi}), we should get the same result, but it did not,  see
Fig.~\ref{fig-drude-compare} here.  The NEGF results are about 2 to 4 times too large!  It isn't easy to understand why. 
The convergence with respect to $L_z$ is very slow.   It took about $L_z \sim 10^3$ to converge (perhaps to the wrong value). 

To understand the slow decay with respect to $L_z$, we note that in a metal, the wave decays according to a length scale
${\rm Im }\,1/q_z$ where
\begin{equation}
q_z = \sqrt{ \epsilon(\omega) \frac{\omega^2}{c^2} - q_\perp^2}.
\end{equation} 
For low frequency $\omega$, it is limited by $q_\perp$, which is set by the scale of gap distance $d$.  When $q_\perp = 0$ and
$\hbar \omega \sim k_B T$,  it is still order 700\,a.u.  In any case, ${\rm Im}\, 1/q_z$ gives a much longer scale than some 
length scales mentioned earlier.   This may be the reason for the much larger $L_z$ for convergence.  If so, the current numerical
scheme is not feasible. 
  
\section{\label{secIX}Resolving the Casimir puzzle?}
Despite the relation between $P$ and $\Pi$ shown in Eq.~(\ref{eq-PtoPi}), and because of the presence of the boundary, it is more convenient to reformulate fluctuational electrodynamics in Coulomb gauge in terms of $P$ and $\Pi_{T}$.  A deeper reason perhaps is that $P$ is local near 
the surface.   This gives as a pair of Langevin-like equations \cite{wang_transport_2023}:
\begin{align}
\label{eq-langevin-scalar}
-\epsilon_0 \nabla^2 \phi &= \rho + P \phi,\\
\label{eq-langevin-vector}
\epsilon_0 (\omega^2 + c^2 \nabla^2) {\bf A} &= -{\bf j}_T + \Pi_T {\bf A}, 
\end{align}
where $\nabla \cdot {\bf A} = 0$, $P \phi$ gives the induced charge and $-\Pi_T {\bf A}$ gives the induced current, 
$\rho$ and ${\bf j}_T$ are the fluctuating charge density and transverse current, which are unconstrained random variables not related
to charge conservation.  The charge conservation only relates $\rho$ to the longitudinal current, ${\bf j}_L$.  These random
fluctuations follow the fluctuation-dissipation theorems,
\begin{align}
\frac{1}{i\hbar} \langle \rho \rho \rangle_\omega & = N\, (P - P^\dagger), 
\\
\frac{1}{i\hbar} \langle {\bf j}_T {\bf j}_T \rangle_\omega &= N\,(\Pi_T - \Pi_T^\dagger). 
\end{align}
 Here $P$ is a scalar (time) and $\Pi$  is 3 by 3 tensor (space), measured per unit volume; $N=1/(e^{\beta \hbar\omega}-1)$ is the Bose function, and the left-hand side is meant to be the correlation in the time domain, Fourier transformed in the frequency domain.  
There could also be cross-correlation terms,
$\langle  {\bf j}_T \rho \rangle$ or $\langle  \rho {\bf j}_T \rangle$, to couple the two equations; but we assume it is negligible.   
The transverse version of $\Pi$ is related to the full $\Pi$ by a projection in $\bf q$ space, as 
\begin{equation}
\Pi_T = (I - \hat{\bf q} \hat{\bf q}) \cdot  \Pi \cdot (I - \hat{\bf q} \hat{\bf q}),  \quad \hat{\bf q} = {\bf q}/|{\bf q}|.
\end{equation}
Here $I$ is the identity tensor, and $\hat{\bf q}$ is the unit vector.    For systems in which $P$ is local, it is 
a good formulation, but this makes the transverse part rather difficult since it is generally non-local.  In particular,
a local Drude model in this framework becomes nonlocal.   So, we gain some
and lose some. 

The Coulomb gauge equations, (\ref{eq-langevin-scalar}) and (\ref{eq-langevin-vector}), can be derived from the
fluctuational electrodynamics Helmholtz equation, $\epsilon_0 (\omega^2 - c^2 \nabla \times \nabla \times  ) {\bf A} = -{\bf j} + \Pi {\bf A}$,
here ${\bf A} = {\bf A}_T +{\bf A}_L$ is the full $\bf A$,  by applying the longitudinal and transverse projectors to this equation, and
rewriting $A_L$ in terms of $\phi$, and ignoring the cross terms; we obtain the pair of equations in the Coulomb gauge. 

The scattering solution associated with Eq.~(\ref{eq-langevin-scalar}) is already given by Eq.~(\ref{r-scalar}) earlier. 
Indeed, the scattering problem of a metal block at $z>0$ in a continuum formulation can be written as
\begin{equation}
-\epsilon_0 \left( -q^2_\perp + \frac{\partial^2 \ }{\partial z^2} \right) \phi = P\, \phi,
\end{equation}
where $P = P({\bf q}_\perp, \omega)$ depends on the transverse wavevector and frequency, local in $z$.  
The ``wave" decays, i.e.,
always in the evanescent modes, we try the scattering solution of the form
\begin{equation}
\phi(z) = \left\{  \begin{array}{ll}
e^{-q_\perp z} + r e^{q_\perp z}, & z < 0, \\
t e^{- k z},  & z \ge 0.
\end{array} 
\right.
\end{equation} 
Here $k = \sqrt{q^2_\perp -P/\epsilon_0}$. 
Matching the boundary conditions at $z=0$ by $\phi(z)$ being continuous and $d\phi(z)/dz$ continuous, we find 
\begin{align}
1 + r & = t, \\
q_\perp (1-r) &= kt.
\end{align}
Solving for $r$, we obtain (designated as now $r_c$ as it is associated with the Coulomb field)
\begin{equation}
r_c = \frac{q_\perp-k}{q_\perp+k},
\end{equation}
which is Eq.~(\ref{r-scalar}).
 
We only need to consider the reflection coefficient $r^T_s$ and $r^T_p$ associated with Eq.~(\ref{eq-langevin-vector}).
Since this is formally the same as the usual problem, except now $\Pi$ is replaced by the transverse version
$\Pi_T$ --- the projection onto the direction perpendicular to $\bf q$, and the $-c^2 \nabla(\nabla \cdot {\bf A})$ term is missing
as $\bf A$ is transverse (the usual equation was $-c^2 \nabla \times ( \nabla \times{\bf  A})$ and the current was the
full current ${\bf j} = {\bf j}_T + {\bf j}_L$). 

If the missing $\nabla \cdot {\bf A}$ does not produce trouble, then our first guess is that the Fresnel coefficients remain
the same as usual, except we need to replace $\epsilon$ by the tensor
\begin{equation}
\epsilon_T =  (I - \hat{\bf q} \hat{\bf q})  \epsilon.
\end{equation}
Here, we assume the original (still Drude) $\epsilon$ is isotropic and proportional to the identity; the transverse dielectric function
simply acquires the transverse projector.   But this is true only for the $s$ polarization.
Speculative, but if we assume the Lifshitz formula, Eq.(\ref{eq-thermal-Casimir}), still holds with three modes:  scalar potential, and two from the transverse vector fields, $s$ and $p$ polarizations in Coulomb gauge,  we just need one of the transverse field reflection coefficients, $r_s^T$ or $r_p^T$, equal $1$ or $-1$.   
The total contribution is then 2 since $r_c \approx -1$, just like what happens if the plasma model were used. 

If we focus on $q_z \approx 0$ because that is important at large distance $d$, we can consider the projector 
a unit vector in the $(x,y)$ sector and zero elsewhere.  Then, it looks the problem is just we have $\Pi_{xx} = \Pi_{yy}$ with
$\Pi_{zz}=0$.  This is what happens numerically for small $\omega$.   It is somewhat more like a 2D plane.  
Because of the transverse projection, $\Pi_{zz} = 0$ has little consequence.  

Here, we might need to work with
\begin{align}
\epsilon_{||}  = 1 - \frac{P}{\epsilon_0 q^2}, \quad \epsilon_T = I - \frac{\Pi_T}{\epsilon_0 \omega^2},
\end{align}
if we eliminate $\phi$ in favor of a full $\bf A$. 

To solve the scattering problem (by setting ${\bf j}_T = {\bf 0}$), we need to consider the differential equation 
\begin{align}
\epsilon_0 \left( \omega^2 - c^2 q_\perp^2 + c^2 \frac{\partial^2\ }{\partial z^2}\right)  {\bf A} = \Pi_T {\bf A} \qquad\qquad\qquad \nonumber   \\
\qquad =  -\epsilon_0 \omega^2 \left(I - \frac{{\bf q} {\bf q}}{q^2}\right)  (\epsilon(z)-1) \left(I - \frac{{\bf q} {\bf q}}{q^2}\right) {\bf A}. 
\label{eq-A-transverse}
\end{align}
Here, $\epsilon(z) = 1$ for $z<0$, and is assumed to be equal to the Drude model value for $z>0$.  There is a discontinuity at 
$z=0$.   The transverse projector on the left and right of $\epsilon(z)$ is constructed with the wavevector $\bf q$, which is a differential operator
\begin{equation}
{\bf q} =  \left( \begin{array}{c}
q_x \\
q_y \\
 - i\frac{\partial}{\partial z}
\end{array} \right).
\end{equation}
and $q^2 = q_\perp^2 + q_z^2 = q_x^2 + q_y^2 - \partial^2/\partial z^2$, also a differential operator in the denominator (this 
is the Coulomb integral kernel, Eq.~(\ref{eq-coulomb-kernel})\,). 

How to solve this equation?  We recall some basic facts in the Coulomb gauge:  the electric field and magnetic induction 
are given by ${\bf E}_T = -\partial {\bf A}/{\partial t}$ and ${\bf B} = \nabla \times {\bf A}$, respectively.   This is so whether 
we have media or not.  So, in frequency domain ${\bf E}_T$ is the same as $i\omega {\bf A}$, and longitudinal field ${\bf E}_L = -\nabla \phi$.  The Gauss law is $\nabla \cdot {\bf E}_L = \rho/\epsilon_0$. The vector potential is transverse, 
so $\nabla \cdot {\bf A} = 0$, for all $z$ (left side, right side, and across the surface).  The Faraday law and magnetic field
Gauss law remains the
same $\nabla \times {\bf E} = \nabla \times {\bf E}_T = i\omega {\bf B}$ and $\nabla \cdot {\bf B} = 0$.  Equation~(\ref{eq-A-transverse})
is equivalent to Ampere's law. 

Let us assume we have a waveform  $e^{\pm i k_z z}$, then we can replace the
operators by the number 
\begin{equation}
k_z = \pm \sqrt{\frac{\omega^2}{c^2} - q_\perp^2}, \quad {\rm Im}\, k_z > 0,
\end{equation}
for $z<0$, and 
\begin{equation} 
k'_z = \sqrt{\epsilon \frac{\omega^2}{c^2} - q_\perp^2},\quad {\rm Im}\, k'_z > 0,
\end{equation} 
for $z>0$, assuming the waveform of $e^{ i k'_z z}$. 
Due to the rotational symmetry of the problem, without loss of generality, we assume
${\bf q} = (q_x, 0, k'_z)$ laying in the $(x,z)$ plane.  We can see that the equations decouple between
$A_y$ and $(A_x,A_z)$.   It is the usual $s$ (TE) and $p$ (TM) polarization designation. 
Due to the transverse condition, $\nabla \cdot {\bf A} = 0$,  $A_x$ and $A_z$ are
related by $iq_x A_x + \partial A_z  /\partial z= 0$. The complicated Eq.~(\ref{eq-A-transverse}) reduces to
three 1D wave equations for each direction $\alpha$,
\begin{equation}
\left[ \epsilon(z) \left( \frac{\omega}{c}\right)^2 - q_\perp^2 + \frac{\partial^2 \ }{\partial z^2} \right] A_\alpha = 0,\quad
z \neq 0. 
\end{equation}
Exactly at $z=0$, there are $\delta(z)$ singularities for $x$ and $z$ components, but not for $y$ ($s$ polarization) component. 
We take the scattering solutions as 
\begin{equation}
A_\alpha = c_\alpha \left\{  \begin{array}{ll}
e^{i k_z z} + r e^{ - ik_z z}, & z < 0, \\
t  e^{ik'_z z},  & z \ge 0,
\end{array} 
\right.
\end{equation} 
for $\alpha = x$ or $y$ and 
\begin{equation}
A_z =  \left\{  \begin{array}{ll}
- c_x \frac{q_x}{k_z} \left( e^{i k_z z} - r e^{ - ik_z z}\right) , & z < 0, \\
- c_x \frac{q_x}{k'_z} t\, e^{ik'_z z},  & z \ge 0,
\end{array} 
\right.
\end{equation} 
with the constants $c_y$ and $c_x$ arbitrary (for $s$ and $p$ polarization, respectively), giving two modes instead of three.  

At this point, we need to discuss the $s$ and $p$ polarization separately, as they are different.  Using the conditions that $A_y = (E_T)_y/(i\omega)$  and 
$\partial A_y/\partial z = - B_x$ are continuous across $z=0$,  we have 
\begin{align}
1 + r_s &= t_s,\\
k_z ( 1 - r_s) &= k'_z t_s.
\end{align}
This gives the same result as the full electrodynamics:
\begin{equation}
r_s^T  =  \frac{k_z - k'_z}{k_z + k'_z} \to 0 \quad {\rm when\ } \omega \to 0.
\end{equation}
Here, prime denotes the metal side, and non-prime is the vacuum.
The continuity of $A_y$ is generally true due to Faraday's law; but the continuity of its derivative in $z$ is valid only if there is no
extra surface current.  This is the case by Eq.~(\ref{eq-A-transverse}).   But there is indeed the possibility of a surface
current due to $\langle {\bf j} \rho \rangle$ coupling which we have not explored. 

The $p$ polarization is fixed entirely by the transverse gauge condition, $\nabla \cdot {\bf A} = 0$. 
The gauge condition is already
used to set up the scattering form for $A_x$ and $A_z$.   In addition to the continuity of $A_x$, which is the same as the continuity of
$(E_T)_x$, which gives $1 + r_p = t_p$, the normal direction must be continuous across the interface, giving
\begin{equation}
\frac{1}{k_z} ( 1 - r_p) = \frac{t_p}{k'_z}.
\end{equation}
Together, we find the reflection coefficient,
\begin{equation}
r_p^T  =  - r_s^T =  \frac{k_z' - k_z}{k'_z + k_z} \to 0 \quad {\rm when\ } \omega \to 0.
\end{equation}
There is a discontinuity across the surface for the magnetic field $B_y(0^+) - B_y(0^-) = 
i (\omega/c)^2 (\epsilon -1) A_x(0)/k'_z$.   This is accounted for by the surface current generated by the discontinuity
of the dielectric function.  Somehow, since $r_p^T$ is fixed by the divergenceless condition only, an additional 
surface current should not change it.  
We can check that all the required boundary conditions are satisfied from the transverse 
Maxwell equations.  
As both $r_s^T$ and $r_p^T \to 0$, we again fail to have the plasma model result.   The thermal Casimir force 
at a large distance is contributed only by the evanescent Coulomb interaction in this picture.

\section{Conclusion}
This project started as a numerical approach to Casimir and heat transfer problems with semi-infinite metal blocks.  However,
due to the convergence issue with respect to $L_z$, it is a rather difficult problem.   
As a result, we focus on the property of $\Pi$ and $P$ of the metal
surface.   These linear response functions depend on the frequency $\omega$, transverse wavevector ${\bf q}_\perp$, and distances
$(j,j')$ away from the surface. 
The surface modification to the corresponding bulk materials properties should be important for both the heat
transfer and Casimir forces, as fundamentally, the scattering problem is a surface or interface problem.    On the way, we 
found the Coulomb interaction alone could give a $1/d^4$ scaling, instead of the common belief of $1/d^2$ as predicted 
by the Drude model.   This $1/d^4$ dependence is due to the fact that $P$ in the ${\bf q}_\perp \to 0$ limit and at small $\omega$ 
is the negative density of states of the electrons on the surface.    $P$ turns out very simple in this limit; it is local in $j$
and takes the form $a + i b \omega$.  

For $\Pi$, we investigate how well it can be described as a local Drude model. 
From the numerical data, we can conclude that the surface layers differ from the bulk in about 100 lattice
spaces, i.e., about 20\,nm.   The local Drude model can be considered correct if we are interested in length scales larger than this.
However, the system near the surface is also quite anisotropic.  The transverse directions almost follow the
Drude model even at the first layer, but normal direction needs about 100 layers.   Since these length scales are comparable
to the mean-free path (or determined by it), and also comparable to skin depths or the convergence thickness for
the transport quantities (heat current or force), it casts some doubt on the usual approach based on continuum models. 
We also encounter numerical difficulty and reliability with our lattice approach, but we encourage researchers 
to investigate new methods based on atomic models. 

The reflection coefficients between a vacuum and a material surface in various settings are analyzed with the aim of resolving the Casimir puzzle. 
We pointed to some directions without success.  Perhaps eventually, the resolution must be
that the interface should be handled from a more fundamental point of view with a correct nonlocal $\Pi$ and $P$ for the surface
taking into account still unknown mechanisms.

\begin{acknowledgments}
The author thanks Pablo Rodriguez-Lopez and Mauro Antezza for many hours of discussions and email exchanges, without
which the paper will not be in its current form. 
J.-S.W. acknowledges support from MOE FRC tier 1 grant A-8000990-00-00.
\end{acknowledgments}

\bibliography{metalcasimir-references}


\end{document}